\documentclass[twocolumn,showpacs,amsmath,amssymb,pre,aps]{revtex4}   
\usepackage{graphicx}
\usepackage{dcolumn}
\usepackage{bm}
\usepackage{color}

\renewcommand{\vec}[1]{\mathbf{#1}}

\newif\ifgraph

\graphtrue             %

\begin{document}
\title{Escape Kinetics of Self-Propelled Janus Particles from a
Cavity: Numerical Simulations\footnote[1]{This paper is dedicated
to Professor Deb Shankar Ray on the occasion of his sixtieth
birthday.}}

\author{Pulak Kumar Ghosh\footnote[2]{Email: pulak.chem@presiuniv.ac.in}}

\affiliation{Department of Chemistry, Presidency University,
Kolkata - 700073, India}


\date{\today}

\begin{abstract}
We numerically investigate the escape kinetics of elliptic Janus
particles from narrow two-dimensional cavities with reflecting
walls. The self-propulsion velocity of the Janus particle is
directed along either their major (prolate) or  minor axis
(oblate). We show that the mean exit time is very sensitive to the
cavity geometry, particle shape and self-propulsion strength. The
mean exit time is found to be a minimum when the self-propulsion
length is equal to the cavity size. We also find the optimum mean
escape time as a function of the self-propulsion velocity,
translational diffusion, and particle shape. Thus, effective
transport control mechanisms for Janus particles in a channel can
be implemented.
\end{abstract}
  \maketitle

\begin{figure}
\centering
\includegraphics[width=0.40\textwidth,height=0.20\textwidth]{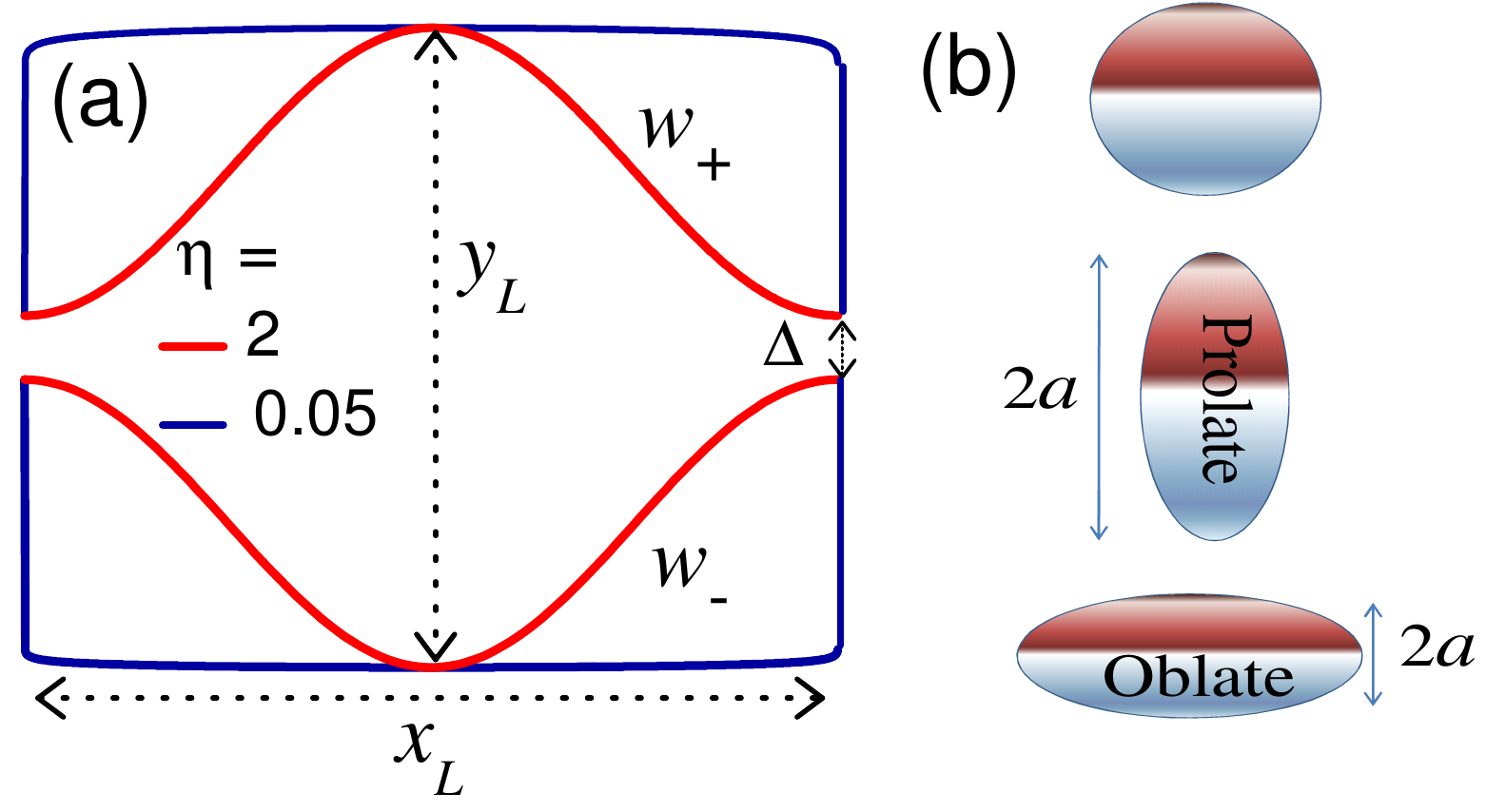}
\includegraphics[height=0.18\textwidth,width=0.45\textwidth]{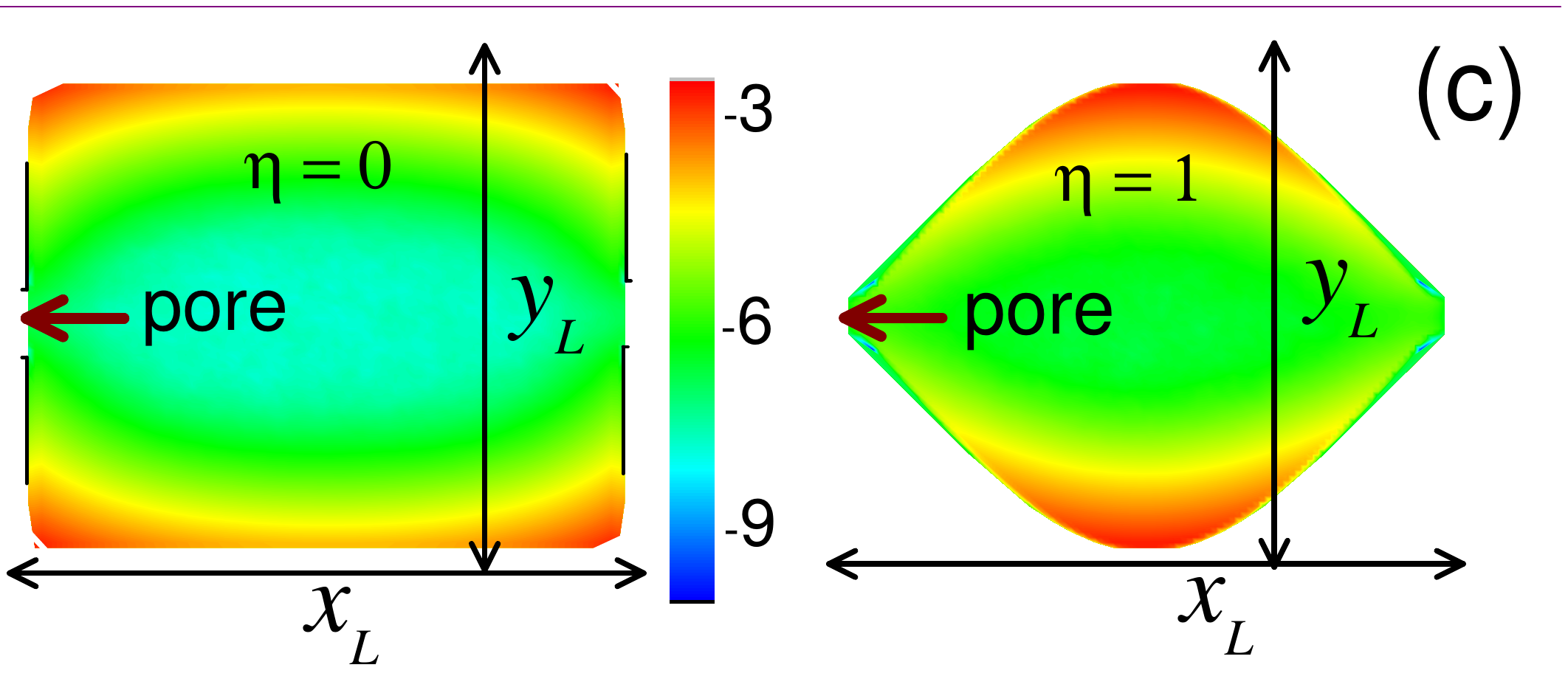}
\caption {(Color online) (a) Shape of the cavity for two different
values of $\eta$ using the wall profile functions, $w_{\pm}(x)$ of
Eq.~(1). (b) Janus particles with various shapes. (c) Logarithmic
contour plots of the stationary particle density $P(x,y)$ in the
cavities. Simulation parameters are: $x_L = y_L = 1,\; \Delta =
0.16, a=b=0.052,\; D_0 = 0.03,\; D_\theta = 0.0005,\; v_0 = 1 $.}
\end{figure}

Self-propelled Janus particles (JPs) are a class of artificial
microswimmers which can move by extracting energy from their
suspension medium \cite{review}. This type of particles consists
of two distinct faces with different chemical or physical
properties. Such two-faced particles can acquire self-propulsion
by inducing chemical concentration or temperature gradients in the
vicinity of active face. A number of
works\cite{cataly1,cataly2,cataly3} show that a controllable
concentration gradient can be created in some catalytic reactions
on the one surface of JPs. Inhomogeneous light
absorption\cite{ther} or magnetic excitation\cite{mag} of JPs can
generate enough local temperature gradient for
self-thermophoresis. Based on different self-phoretic mechanisms
various kinds of self-propeller have been designed with specific
goals\cite{review,sano2}

In the absence of any external force field, the motion of a
self-propelled JP is directed parallel to the self-phoretic force.
Gradient fluctuations or collisions with boundaries or the
intrinsic rotational diffusion result in a random change of the
direction of self-propulsion. Thus, self-propelled JPs exhibit
time correlated active Brownian motions. Janus particles can be
used as a special kind of diffusing tracer in experiments aimed at
demonstrating  non-equilibrium phenomena like,
ratcheting\cite{MSshort,ratchet}, autonomous pumps\cite{MSshort},
absolute negative mobility\cite{GNM} etc. Moreover, it would be
desirable to gain control over the motion of this class of
Brownian tracers, so as to use them as a ``nano-robot" for
applications in the medical sciences and nano-technology.

A conspicuous feature reported in earlier
experiments\cite{Bechinger} and simulations\cite{MSshort} is that
when the mean free path ($l_\theta $) of a JP is much greater than
the cavity size ($x_L \; {\rm or} \; y_L$), the  particle spends
most of their time in the close vicinity of confining walls. In
many practical situations, the mean free path\cite{MSshort}
follows the condition $l_\theta \gg$ {$x_L \; {\rm or} \; y_L$}.
The self-propulsive forces press JPs against walls. As a result,
the JPs keep diffusing in the tangential direction under the
action of translational noises until an appropriate orientational
change occurs by rotational diffusion. Taking advantage of this
property, JPs can be captured by placing an obstruction of
appropriate shape \cite{capture},  driven against an applied
force\cite{GNM}, and
 rectified in asymmetric channels with high efficiency\cite{MSshort}. All these features make the
dynamics of
 JPs different from living micro-swimmers\cite{run1,run2,run3}
(\emph{e.g.}, Bacteria). Living micro-swimmers change their
direction whenever they encounter any obstruction on their path
(run-and-tumble like dynamics\cite{run2}).

In this paper we explore how the escape kinetics of a JP out of a
cavity can be controlled by tuning the self-propulsive properties
as well as the shape of both, the particle and confining walls. In
confined systems, the boundary conditions (which are determined by
the shape of particles and confining walls) govern the Brownian
dynamics\cite{ChemPhysChem,pkg2,borro}. Therefore, transport
control of a JP with assigned self-propulsive properties and shape
can be achieved only by suitably tailoring the channel boundaries.

On the one hand,  self-propulsion tries to confine JPs at some
corners of the cavity. On the other hand, the translational noises
tend to contrast the action of self-propulsion by broadening the
localized JP densities.  Moreover, effects of self-propulsion
largely depend on the shape of the particle and confinement. Thus,
the interplay among the translational noises, self-propulsion and
particle shape produces a rich JP dynamics.

{\it Model.} --- We consider an elongated  self-propelled JP
diffusing in a two-dimensional (2D) cavity (extension of
conclusions to 3D is straightforward). Elongated JPs  have been
modelled as elliptical disks with major and minor axes $2a $ and
$2b $, respectively. There are a number of well-established
methods to synthesize such elongated JPs\cite{review,cataly1}. The
walls of the cavity have been modelled by the following sinusoidal
functions [depicted in Fig.~1(a)]
\begin{eqnarray}
\label{wx} w_{\pm}(x) = \pm \frac{1}{2} \left [\Delta
+(y_L-\Delta)\sin^{\eta} \left(\frac{\pi x}{x_L}\right) \right],
\end{eqnarray}
where $x_L$ and $y_L$ are the length and width of the cavity.
$\Delta$ is the pore size through which the JPs can exit the
cavity. An additional tunable geometric parameter $\eta$ has been
introduced to reproduce most of the cavity geometries investigated
in the literature\cite{ChemPhysChem,borro,ai,pkg1,Debasish}. For
$\eta = 2$, the cavity represents the compartment of sinusoidally
corrugated channel \cite{pkg2,ChemPhysChem}. Moreover, when $\eta
\rightarrow 0$, the cavity reproduces the compartment of sharply
corrugated channels, where geometric effects are much more
prominent than the former case \cite{borro}.  The bulk dynamics of
a self-propelled JP can be described by the following
equations\cite{Lowen},
\begin{eqnarray}
 \dot x = v_0\cos \theta +\xi_x(t),\;\;\;\;\;
 \dot y =v_0\sin \theta +\xi_y(t),\label{Lan2}
\end{eqnarray}
where $(x,y)$ denote the position of the particle center of mass.
The particle diffuses under the action of self-propulsion and
equilibrium thermal fluctuations. We assume that the
self-propulsion velocity, $\vec{v}_0$, is oriented along either
the major (for prolate) or  minor  (for oblate) axis of the
particle. The vector $\vec{v}_0$ makes an angle $\theta$ with the
$x$-axis of the cavity. Due to rotational diffusion of the
particle, $\theta$ changes randomly, which can be described as a
Wiener process, $\dot \theta=\chi_\theta(t)$, with $\langle
\chi_{\theta}(t)\rangle=0$ and $\langle
\chi_{\theta}(t)\chi_{\theta}(0)\rangle=2D_\theta\delta (t)$,
where the rotational diffusion constant $D_{\theta}$ is related to
the viscosity ($\eta_v$) of the medium, temperature ($T$) and size
of the particle. For an elliptical particle, $D_{\theta} \propto
k_BT/ab\eta_v$. From the correlation function, $\langle \cos
\theta (t) \cos \theta (0) \rangle = \langle \sin \theta (t)\sin
\theta (0)\rangle =1/2 e^{-|t|D_\theta}$, one can consider the
self-propulsion velocity components, $v_{x}=v_0 \cos \theta$ and
$v_{y}=v_0 \sin \theta$, as the components of a 2D non-Gaussian
noise ${\vec \chi_{c,i}}(t)$ with zero mean, $\langle
\chi_{c,i}(t)\rangle=0$, and finite-time correlation functions,
$\langle
\chi_{c,i}(t)\chi_{c,j}(0)\rangle=2(D_c/\tau_\theta)\delta_{ij}e^{-2|t|/\tau_\theta}$,
where $i=\{x,y\}$, and $D_c=v_0^2\tau_\theta/4$ with
$\tau_\theta=2/D_\theta$ [8]. The last terms of the
Eqs.~(\ref{Lan2}) [$\xi_{x}(t)$ and $\xi_{y}(t)$] are the thermal
noise responsible for translational diffusion of the JP.
$\xi_x(t)$ and $\xi_y(t)$) can be modelled by Gaussian white
noises with $\langle \xi_{i}(t)\rangle=0$ and $\langle
\xi_{i}(t)\xi_{j}(0)\rangle=2D_0\delta_{ij}\delta (t)$, where
$D_0$ is the measure of the translational diffusion of a JP in the
bulk with $v_0 = 0$. In the bulk, the two equations in
Eqs.~(\ref{Lan2}) are statistically independent, so the particle
diffuses according to F\"urth's law \cite{MSshort,Marchetti}.

The mechanisms of the translational and rotational diffusion may
not be the same and therefore $D_0$, $v_0$, and $\tau_\theta$ can
be treated as independent model parameters. Moreover, for the sake
of simplicity, we have ignored particle-particle
collisions\cite{Buttinoni} and hydrodynamic effects\cite{Ripoll}.
Despite all these simplifications, analytical calculations of the
mean exit time out of a cavity is a formidable task. Therefore, we
resort to numerical simulations to accomplish our goals.

\begin{figure}
\centering
\includegraphics[height=0.27\textwidth,width=0.45\textwidth]{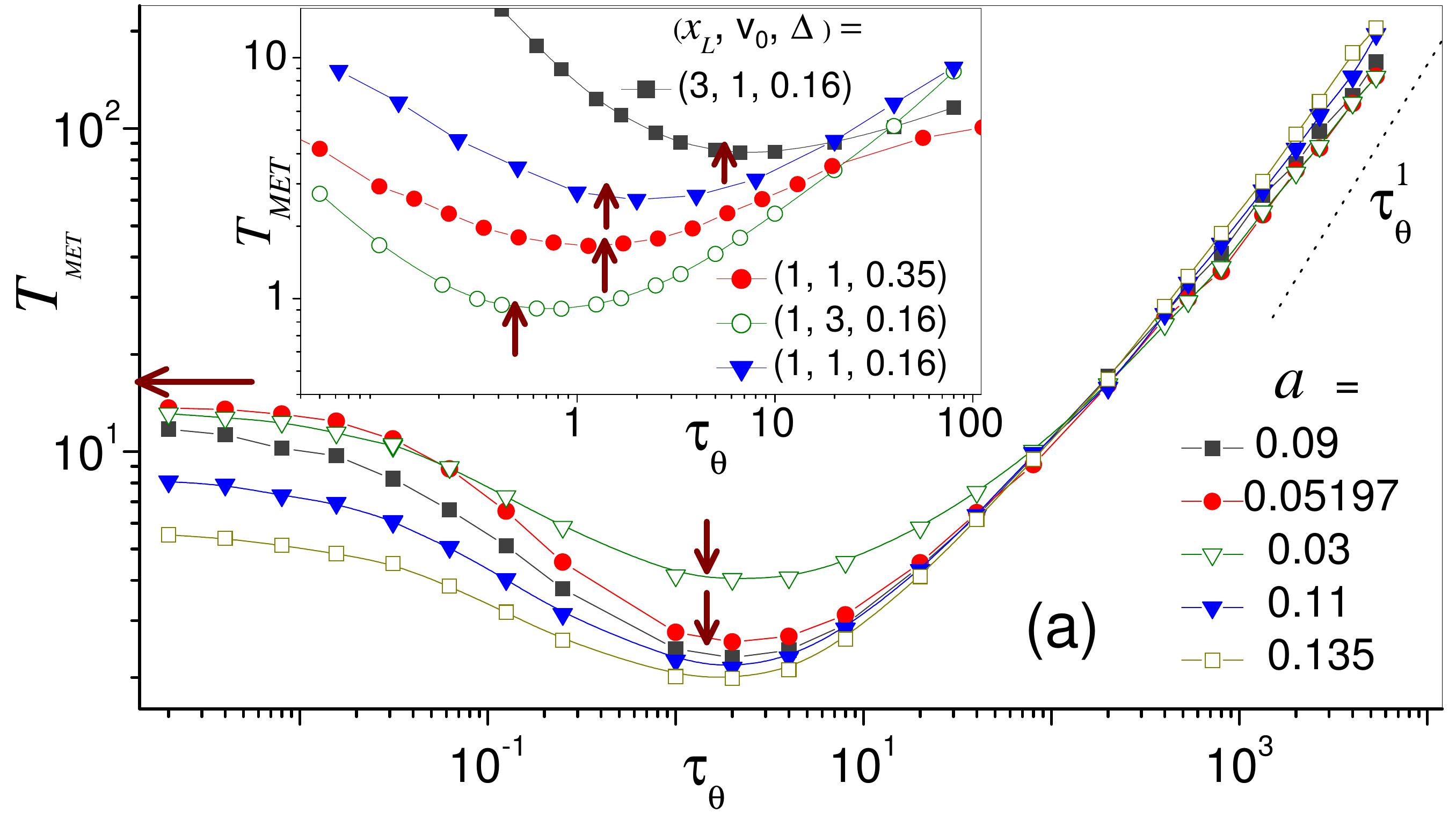}
\includegraphics[height=0.27\textwidth,width=0.45\textwidth]{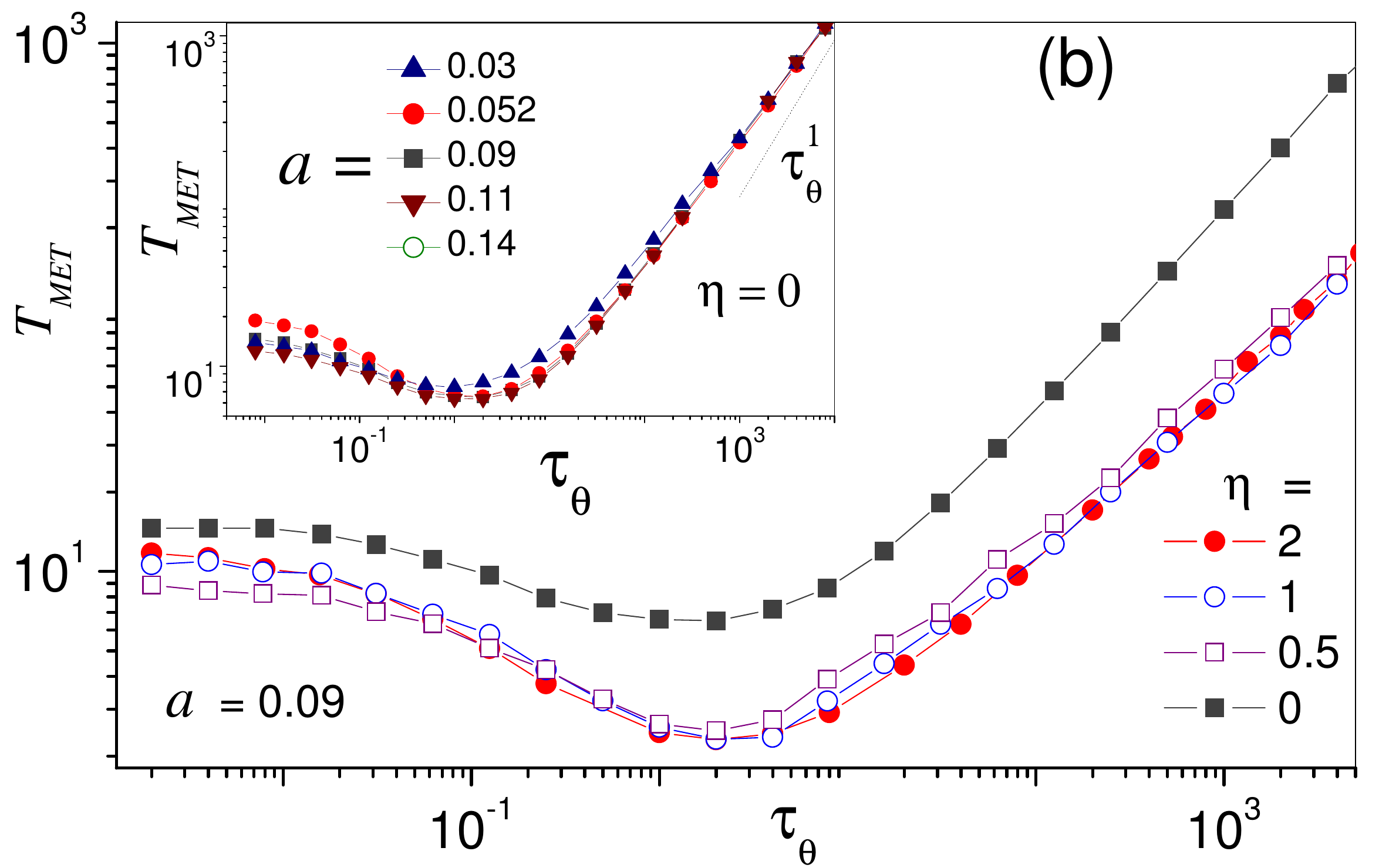}
\caption {(Color online) (a) $T_{MET}$ {\it vs.} $\tau_{\theta}$
for prolate, oblate and circular JPs. Semiaxes $a$ and $b$ are
varied keeping, $ab= 0.0027$. Cavity parameters are:  $x_L=
y_L=1$, $\Delta=0.16$, and $\eta = 2$. Other parameters are:
$v_0=1 $ and $D_0 = 0.03$.  The prediction of Eq.~(\ref{as1}) has
been  displayed by a horizontal arrow. The dotted line is the
power law $\tau_\theta^1$ drawn for the reader's convenience.
Inset present variations of $(\tau_\theta)_m$ in $T_{ {\rm MET}}$
\emph{vs.} $\tau_\theta$ plots, with $v_0$, $x_L$ and $\Delta$
(see legends). Vertical arrows denote position of minima based on
the Eq.~(\ref{min}).  Parameters are the same as the main figure
but $a=0.052$.
\newline (b) $T_{ {\rm MET}}$ {\it vs.}
$\tau_{\theta}$ for different cavity shapes. The inset depicts the
same for $\eta = 0$ and different particle shapes. Parameters are
the same as Fig.~2(a) except those are mentioned in the legends.}
\end{figure}

We numerically estimate the mean exit time ($T_{ {\rm MET}}$) of
JPs out of the compartment of corrugated channel. The mean exit
time is defined as the average time a JP requires to exit a
channel compartment starting from a random initial position
($x,y$)  and orientation ($\theta$) within the cavity. The Eqs~(2)
have been numerically integrated under the assumption that the
channel walls are perfectly reflecting and the particle-wall
collisions are elastic \cite{ANMshape}. All the results (presented
in the Figs.~2-3) are obtained by ensemble averaging over $10^4 -
10^6$ trajectories depending upon the values of parameters. We
choose times in seconds and lengths in microns (see
reference\cite{unit}).

{\it Mean exit time versus rotational diffusion.} --- To
understand the underlying escape mechanisms of a JP out of a
cavity, we first explore $T_{ {\rm MET}}$ as a function of the
rotational time constant $\tau_\theta$ (shown in Fig.~2). The
rotational time constant is defined as the average time during
which a swimming JP maintains its direction away from the walls.
$\tau_\theta$ is related to the bulk rotational diffusion constant
as $\tau_\theta = 2/D_\theta $. Figure~2 shows that for
$\tau_\theta \rightarrow 0$, the mean exit time attains a constant
value. In the opposite limit, when $\tau_\theta \rightarrow \infty
$ the escape rate is
 approximately inversely proportional to $\tau_\theta $. Between
 these two limits, a minimum is observed in the $T_{ {\rm MET}}$
\emph{vs.} $\tau_\theta$ plot. All these features can be explained
by the following considerations.
\newline (i) For $\tau_\theta \rightarrow 0$, the self-propulsion velocity
changes its direction instantaneously. Thus, the propulsive force
acts as a zero mean white noise and its contribution to diffusion
is negligibly small as $D_0 \gg \tau_\theta v_0^2 /4$. In this
regime, the $T_{\rm MET}$ of a spherical JP can be calculated
analytically using the Zwanzig-Fick-Jacobs scheme  for entropic
channels \cite{ChemPhysChem,FJ,Zwanzig} or the random walker
scheme\cite{bosi,Dagdug}.
\begin{eqnarray}
T_{MET} =
\frac{x_L^2}{8D_0}\sqrt{\frac{y_L}{\tilde{\Delta}}}\left[1+\frac{\tilde{\Delta}}{y_L}
 \right]  \label{as1}
\end{eqnarray}
where $\tilde{\Delta} = \Delta - 2a$ is the effective pore size.
Our simulation results are in good agreement with the predictions
of Eq.~(\ref{as1}) [indicated by an horizontal arrow in
Fig.~2(a)]. However, this estimate is not valid for rod-shaped
particles. Prolate JPs glide along the boundaries so that they can
go through the the pore without any change of their orientation.
Thus, prolate-shaped JPs take less time  to exit in comparison to
circular or oblate JPs.
 \;
\newline (ii) For $\tau_\theta \rightarrow \infty$, the rotation of
JPs against the self-propulsion is the bottleneck of the problem.
To escape from the cavity, a JP requires some orientational
changes for two reasons. Firstly, to be free from the sharp
corners or lobes of the cavity where particles may get stuck, and
secondly to be aligned to the axis of the pore of the cavity. For
the parameter set of Fig.~2, the former one is the rate
determining step when $D_\theta \rightarrow 0$. The numerical
simulation of the stationary particle density $P(x,y)$  [shown in
Fig.~1(c)] corroborates this assertion. For $\tau_{\theta}
\rightarrow \infty$, a power law, $T_{ {\rm MET}}$ =
$A\tau_{\theta}^\alpha$ can be fitted to the simulation data. The
pre-factor $A$ and the exponent $\alpha$ are independent of the
shape of the particle. But $A$ depends on the geometry of the
cavity. To exit from the cavity, if a JP needs to rotate to an
angle $\theta_1$ starting from a randomly chosen angle in between
$\theta_1$ and $\theta_2$, the average escape time can be
estimated as\cite{borro},
\begin{equation}
T_{MET} \sim (\theta_2 - \theta_1)^2/6D_\theta =
\tau_\theta(\theta_2 - \theta_1)^2 /12 \label{as2}
\end{equation}
However, our simulation results (see Fig.~2) show that $\alpha
\sim 0.9$. This mismatch is due to the fact that the derivation of
Eq.~(\ref{as2}) tacitly assumes \emph{free} rotational diffusion
of JPs. But this is not the case practically. Self-propulsion
pushes the JPs against the walls which can enhance the rotational
diffusion due to the smooth curvatures of the confining walls.
Even for $\eta=0$ the excluded-volume considerably reduces the
effects of the sharp corners in the escape kinetics.
\begin{figure}
\centering
\includegraphics[height=0.27\textwidth,width=0.45\textwidth]{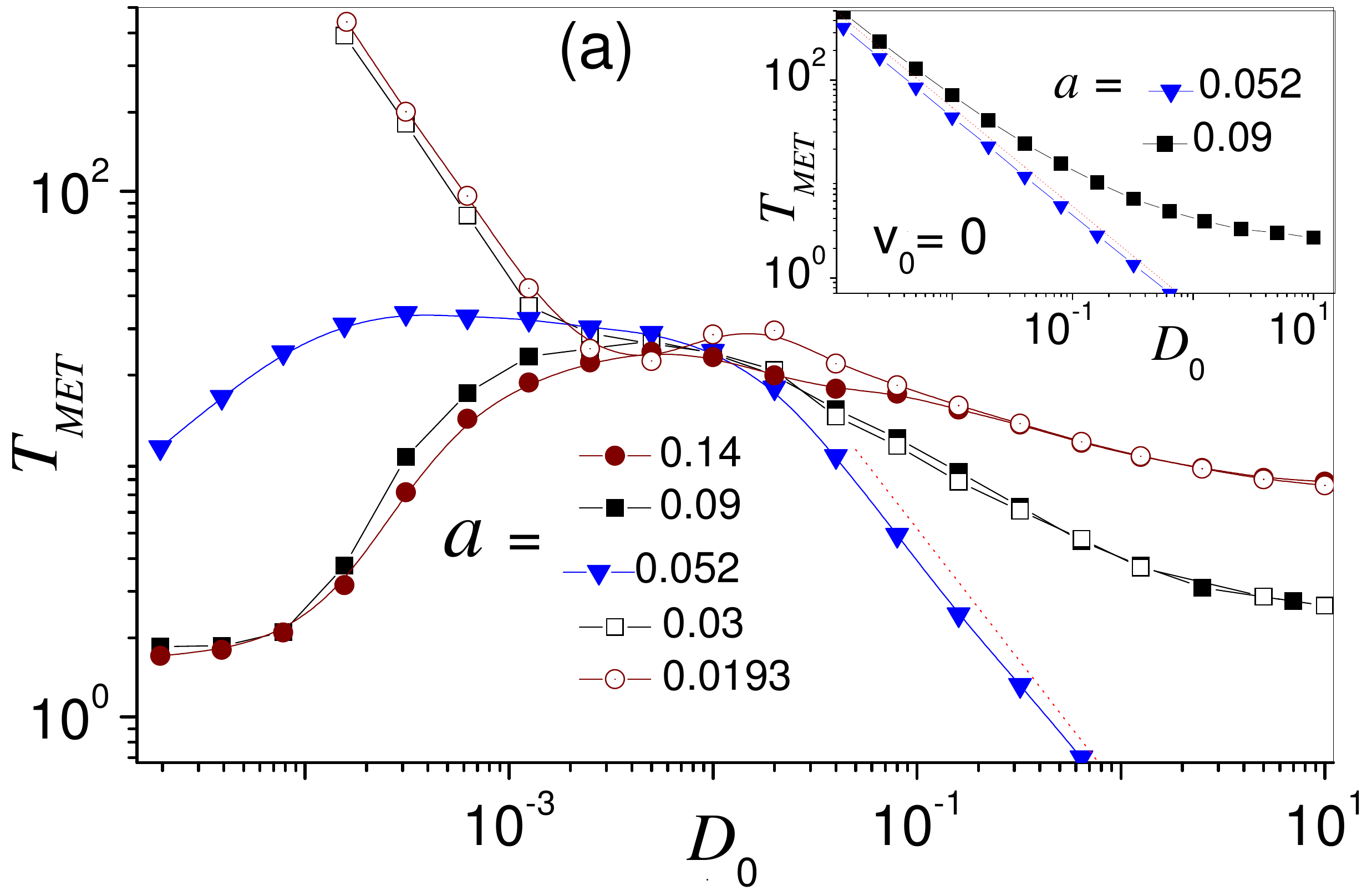}
\includegraphics[height=0.32\textwidth,width=0.45\textwidth]{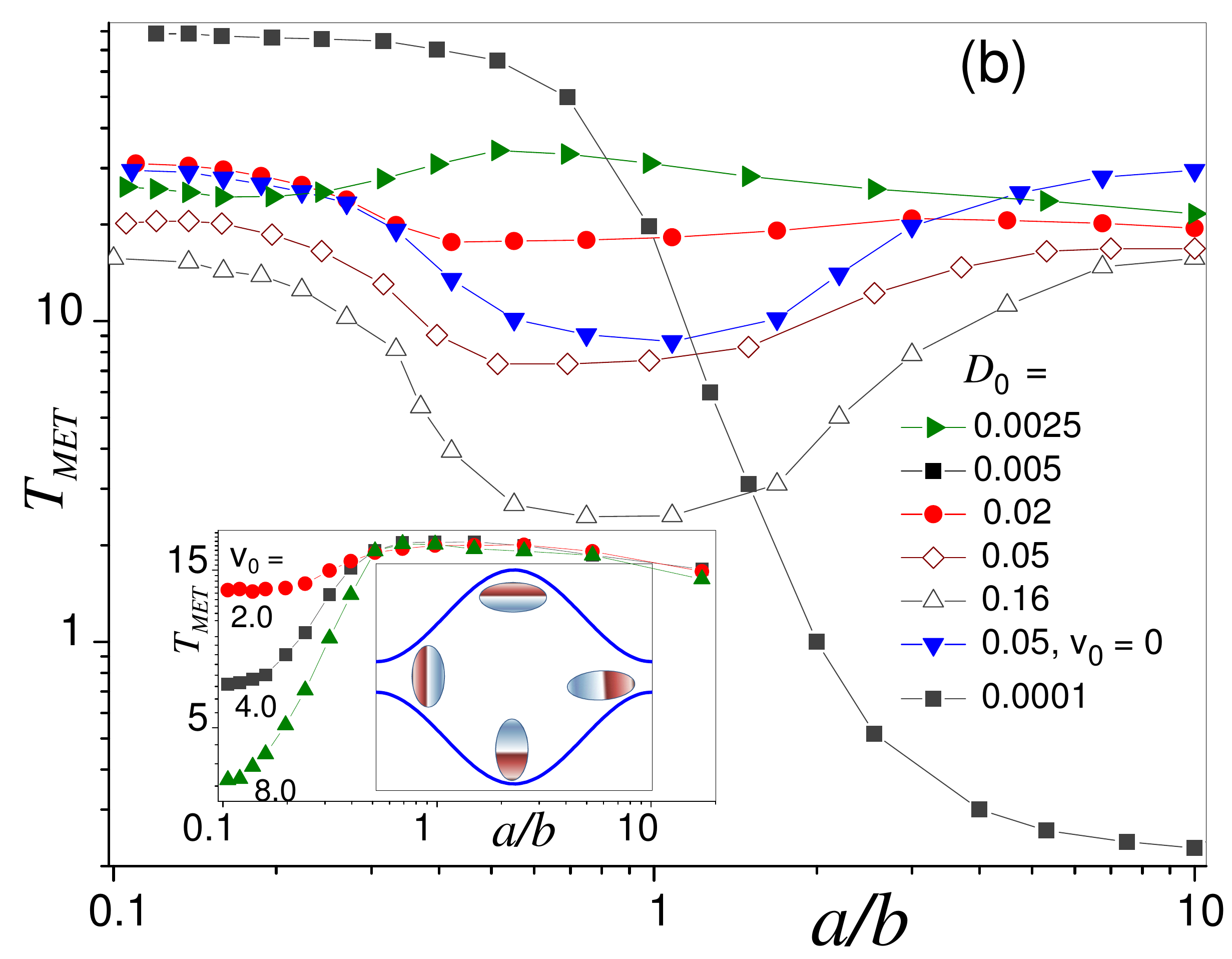}
\caption {(Color online) (a) $T_{ {\rm MET}}$ {\it vs.} $D_0$ for
prolate, oblate and circular JPs. Compartment parameters:
$\eta=2,\;x_L=y_L=1$ and $\Delta=0.16$; self-propulsion
parameters: $v_0=0.5$ and $D_\theta=0.01$; particle size:
$ab=0.0027$. Parameters for the inset is the same as the main
figure, but $v_0$ = 0. The red-dotted lines are the fitting law of
Eq.~(\ref{as1}).
 \newline (b) $T_{{\rm MET}}$ {\it vs.} $a/b$ for the different translational noise
strengths. For all the curves $v_0 = 0.5$ unless it is mentioned
in the legend. Remaining parameters are same as subfigure (a).
$T_{ {\rm MET}}$ has been multiplied by $0.1$ for the curve with
$D_0 = 10^{-4}$ to bring all the curves in the same frame. Inset:
$T_{ {\rm MET}}$ {\it vs.} $a/b$ for different strength of
self-propulsions (see legends). Other parameters are same as
Fig.3(a). Sketch: Some most probable orientation of elongated JPs
with which they approach to the opening and walls in lobes.}
\end{figure}
\newline (iii) In the intermediate regime, on
 increasing $\tau_\theta$ the mean exit time first decreases,
then increases passing through a minimum. The orientational angle
$\theta$ of a JP may have any value between $0$ to $2\pi$. On
average, JPs drift toward the left or right exit with a velocity
$\bar{v} = v_0\cos{\pi/4}$. When the mean free path $l_\theta^x =
\tau_\theta \bar{v}$ along the cavity axis is equal to the length
of the cavity, a large fraction of the trajectories can reach the
boundary (also in the vicinity of the pores) without any
orientational change. Moreover, $\tau_\theta$ is not  large enough
to keep waiting a JP long to acquire the favorable orientations to
exit from the cavity. Thus, $T_{ {\rm MET}}$ \emph{vs.}
$\tau_\theta$ plots exhibit a minimum at,
\begin{equation}
(\tau_\theta)_{m} = \sqrt{2}\;x_L/v_0 \label{min}
\end{equation}
It is apparent from Fig.~2 that the numerical analyses match
fairly well with the above estimates. Figure~2(a) clearly
indicates that $(\tau_\theta)_{m}$ is independent of the shape of
the JPs. Figure 2(b) shows that the dependence of $T_{ {\rm MET}}$
on $\tau_\theta$ is independent of the cavity shape. However, for
$\eta = 0$ the cavity possesses some sharp corners where JPs may
get stuck. Becoming free from this type of stuck states is more
difficult than the stuck state in a lobe of sinusoidal channel
compartments. Thus, a JP takes less time to cross the compartment
of smoothly corrugated channel than a sharp one.

{\it Mean exit time versus translational noises}
--- Figure 3(a) depicts $T_{ {\rm MET}}$ as a function of the strength
of translational noise $D_0$. When $D_0$ is very small  in
comparison to $v_0$, the escape kinetics of the prolate JPs is
solely guided by self-propulsion\cite{dimension}. After a
collision with the wall, the prolate swimmers tend to slide
parallel to the wall and get out of the cavity whenever they find
an opening. These swimmers slide on the walls in such a way that
they do not need any further orientational change to exit the
cavity. As a result, for $D_0 \ll v_0$, the mean exit time is
independent of $D_0$ for prolate or circular JPs. The oblate
swimmers too tend to pile up against the wall and try to slide,
but, due to their shape the longitudinal diffusion is
suppressed\cite{GNM}. Moreover, the oblate particles need
assistance by thermal fluctuations $\xi_i$ to get aligned so as to
escape. This leads to a suppression of the exit rate of oblate JPs
with decreasing $D_0$. Upon increasing the intensity of $\xi_i$,
noises start kicking the particle out of its sliding or stuck
states. As a result, the escape process of oblate particles is
facilitated, whereas the exit process of prolate and circular
swimmers gets retarded. When $D_0 \gg v_0$ the translational
diffusion owing to $\xi_i$ dominates over the propulsive force. In
this regime, the escape rate of a circular JP is given by
Eq.~(\ref{as1}). But elongated particles cannot go through the
pore unless they acquire some specific orientational angles by
rotational diffusion. Thus, the exit rate of prolate or oblate JPs
becomes insensitive to $D_0$ for $D_0 \rightarrow \infty$ and the
asymptote can be determined by making use of Eq.~(\ref{as2}).


{\it Mean exit time versus particle shapes} --- To better
understand the interplay between the translational noise strength
and the particle shape in the escape kinetics, we estimate $T_{
{\rm MET}}$ as a function of $a/b$ keeping $ab=$ fixed [see
Fig~3(b)]. For large $D_0$ or in the absence of self-propulsion
the $T_{ {\rm MET}}$ \emph{vs.} $a/b$ plots take a symmetric
U-shape with two horizontal asymptotes. In this regime, $T_{ {\rm
MET}}$ has a minimum for $a=b$ (circular JPs) and maximum for
$a\gg b$ or $a\ll b$ (elongated JPs). Upon decreasing the noise
intensity $D_0$ the curves in Fig.~3(b) become asymmetric and the
exit time for the prolate JPs become different from the oblate
ones. This result can be understood based on the arguments of the
preceding paragraphs. The most surprising result is that circular
particles take the longest time to escape from the cavity,  while,
in certain $D_0$ regimes,  the oblate ones take the shortest. This
attributes to the trapping of the particles at the lobes of the
cavity [see sketch in the inset of Fig.~3(b)]. The sticking
mechanism and the stuck states have some interesting features to
note: \newline (i)  For $D_0 \rightarrow $ 0,  self-propulsion
tends to press the particle against the walls, while, interactions
with walls having a smooth curvature guide the particle to slide
on it. Thus, the particle does not get stuck in the lobes of the
cavity and rotation of the particles against the self-propulsion
is the slowest step. Therefore, the prolate or circular JPs can
exit much faster than the oblate ones. The $T_{ {\rm MET}}$
 \emph{ vs.} $a/b$ plot for $D_0 = 10^{-4}$ in
Fig.~3(b) corroborates this assertion. Also in the opposite limit,
$D_0\gg v_0$  the sticking mechanism has little impact on the
escape dynamics as the noises kick the particle out of its sliding
or stuck states. Between these two limits, there is an
intermediate regime of $D_0$ where the effects of trapping at the
lobes become important in the escape kinetics. \newline (ii) As
anticipated based on geometric considerations regarding the most
probable orientation of the particles near walls [see the sketch
in the inset of Fig.3(b)], a circular disk get stuck in the lobes
most tightly and an oblate one rather weakly. Thus, in the
presence of a very strong self-phoretic force oblate particles can
escape from the cavity much faster than the circular or prolate
ones.

In \emph{conclusion}, depending upon the relative strength of the
propulsion  and the translational noise, two kinds of
noise-activated processes, (a) rotation against the propulsive
force and (b) noise-induced hopping from the trapped states in
lobes or sharp corners, play the central role in the JPs escape
kinetics. When acquiring an appropriate orientation to exit is the
rate determining step, a prolate JP has an order of magnitude
larger escape rate than the particles with other shapes. On the
other hand, when the trapping mechanism dictates the escape
kinetics, oblate JPs require less time to exit from  a cavity in
comparison to prolate or circular ones. Moreover, the escape rate
from a cavity can be maximized by adjusting the self-propulsion
strength in such a way that the mean free path of the JP matches
with the cavity size. The self-propulsion velocity of a JP driven
by chemical reactions can be tuned by changing concentrations of
reactants or catalysts\cite{cataly1,cataly2,cataly3}. Again, in
the experiential set up of light-driven JPs\cite{Bechinger}, one
can easily control the self-propulsion velocity by adjusting the
intensity of light. Therefore, our simulation results can be used
to design most efficient JPs for targeted drug delivery, and many
applications in natural and artificial devices.

Acknowledgments -- I wish to thank Professor F. Marchesoni
(Universita di Camerino, Italy) and Dr. D. P. Chatterjee
(Presidency University, India) for critical comments and
suggestions.


\begin{references}
\bibitem{dimension} from a dimensional point
of view $D_0$ should be compared with $v_0x_L$, which in the
present case equals one.
\end{references}
\end{document}